\documentclass[aps,floatfix,preprint,footinbib]{revtex4}
\usepackage{graphicx,subfigure,amsmath,bbm}

\DeclareMathOperator{\sech}{sech}

\usepackage{chapterbib}

\begin{document}

\title{Controlled transportation of light by light at the microscale}

\author{{ Manuel Crespo-Ballesteros* and Misha Sumetsky}\\
{\small \em Aston Institute of Photonic Technologies, Aston University, Birmingham B4 7ET, UK}\\
{\small *m.crespo@aston.ac.uk}}

\date{\today}

\begin{abstract}

We show how light can be controllably transported by light at microscale dimensions. We design a miniature device which consists of a short segment of an optical fiber coupled to transversally-oriented input-output microfibers. A whispering gallery soliton is launched from the first microfiber into the fiber segment and slowly propagates along its mm-scale length. The soliton loads and unloads optical pulses at designated input-output microfibers. The speed of the soliton and its propagation direction is controlled by the dramatically small, yet feasible to introduce permanently or all-optically, nanoscale variation of the effective fiber radius.

\end{abstract}

\maketitle

Transportation of objects by other objects, both at the macroscale and microscale, is an evident constituent of the evolution of nature in general and living beings in particular. At the macroscale, we travel and carry things from one place to another and use machines to make it easier and faster \cite{Janic2016}. At the microscale, human-guided transportation and manipulation of objects is of great multidisciplinary importance with applications ranging from medical and life science to nanomaterial science, bionanotechnology, and nanoelectronics \cite{Sun2015,Zhang2019,Makulavicius2020}.

In microphotonics, addressed in this paper, we can separate the micro-objects under study into those constituted of matter (e.g., waveguides, microresonators, and micro/nanoparticles) and those constituted of light (e.g., optical waves, pulses, and localized states). Consequently, at the microscale we distinguish the transportation and manipulation of (a) matter by matter, (b) light by matter, (c) matter by light, and (d) light by light.

\begin{figure}
\includegraphics[width=0.75\linewidth]{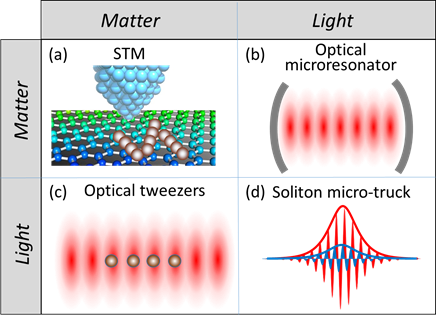}%
\caption{Matter by matter, light by matter, matter by light, and light by light transportation at the microscale. (a) An AFM tip translating atoms at the solid surface. (b) An optical microresonator translating an optical eigenstate. (c) An optical tweezers translating micro-objects. (d) An optical soliton translating an optical eigenstate. \label{fig:fig1}}
\end{figure}

There are numerous examples when matter micro/nano-objects are controllably transported and manipulated by other matter micro/nano-objects. The developed approaches often resemble the manipulation of macroscopic objects with mechanical, electro-mechanical, and magnetic tools \cite{Sun2015,Zhang2019,Makulavicius2020}. In particular, at the atomic-scale dimensions, the transportation and manipulation of atoms and nano-objects are possible with an atomic force microscope (AFM) \cite{Binnig1986,Voigtlander2019,Santos2019} (Fig. \ref{fig:fig1}(a)). 

Transportation of light by material micro-objects is possible as well. For example, optical microresonators are used to confine and manipulate light at the microscale \cite{Vahala2003,Matsko2009} (Fig. \ref{fig:fig1}(b)). They are commonly considered at rest with respect to the laboratory system of reference \cite{Vahala2003,Matsko2009, Foreman2015}.  Generally, the translation of a microresonator with constant speed does not affect the behavior of localized states residing in it \cite{Note1}. However, accelerated translation, vibration and rotation can significantly perturb the resonant states \cite{Carmon2005,Savchenkov2007, Liang2017, Lai2019, Jiang2020}. In the simplest case, light confined in a microresonator can be transported \textit{mechanically }using a ``truck'' in the form of a translation stage. In another example, controlling the perturbation of an eigenstate in a rotating microresonator allows one to use it as a miniature gyroscope \cite{Liang2017, Lai2019}.

In turn, light in the form of optical tweezers can confine and manipulate matter micro-objects \cite{Ashkin1986,Zemanek2019}. For example, light waves can localize microparticles close to their antinodes by the gradient and scattering forces (Fig. \ref{fig:fig1}(c)). In addition, propagation of light through nonlinear media allows the manipulation of light itself, such as modification of its spectrum and self-localization at the microscale. Examples of current significant interest include frequency comb generation \cite{Kippenberg2011, Kippenberg2018a}, optomechanical processes \cite{Aspelmeyer2014} and formation of solitons \cite{Agrawal2012, Kivshar2003, Kartashov2011, Kippenberg2018a}. Conventionally, for telecommunication applications, broadband picosecond and sub-picosecond temporal solitons are used \cite{Agrawal2012, Kivshar2003}. Broadband solitons have been studied in application to the frequency comb generation \cite{Kippenberg2011, Kippenberg2018a}. Slow broadband solitons with picosecond duration have been demonstrated in photonic crystals \cite{Colman2010, Malaguti2012}. \textit{Slow narrowband solitons}, which may have much smaller propagation speed, can be realized in periodic microstructures provided that the soliton central frequency is close to the band gap edge \cite{DeSterke1988, Bhat2001, Mok2006}.

\begin{figure}
\includegraphics[width=0.75\linewidth]{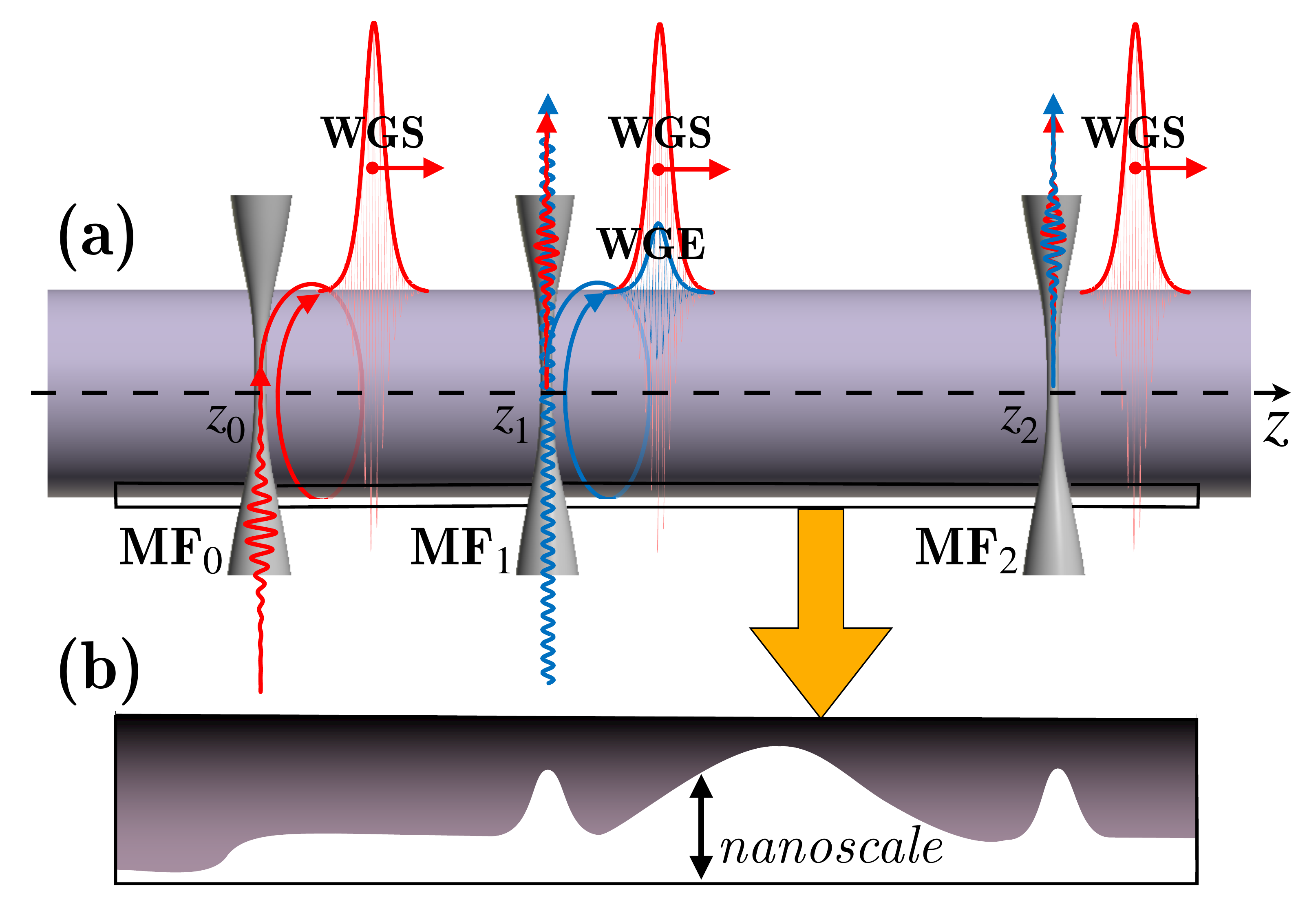}%
\caption{Illustration of the conveyance of a weak optical pulse by a soliton. ${\mathrm{MF}}_0$, ${\mathrm{MF}}_{\mathrm{1}}$, and ${\mathrm{MF}}_{\mathrm{2}}$ are the input-output microfiber waists of biconical tapers coupled to the optical fiber segment. WGS is a whispering gallery soliton, WGE is a relatively weak whispering gallery pulse or an eigenstate transported by the WGS. ${\mathrm{MF}}_{\mathrm{2}}$ serves as the WGS source and ${\mathrm{M}\mathrm{F}}_{\mathrm{1}}$ and ${\mathrm{MF}}_{\mathrm{2}}$ serve as stops where the WGE is loaded and unloaded.} \label{fig:fig2}
\end{figure}

However, can we use light as optical tweezers for light? Is it possible to confine and translate light controllably and \textit{all-optically} at the microscale?  The natural approach to answer this question is to consider \textit{a soliton as a moving microresonator} which can confine and transport weaker light and, thus, be used as a micro-truck for light (Fig. \ref{fig:fig1}(d)). This may be possible since the electric field $E$ of a soliton propagating along an optical fiber induces a change in the refractive index $\Delta n \sim |E|^2$ due to the non-linear Kerr effect [18, 19]. Therefore, the soliton field (as well as the field of any other sufficiently strong optical pulse) can act as a moving effective potential well which traps and transports an optical signal. Three decades ago, transportation of a localized optical state by an optical soliton was proposed \cite{Manassah1990a, Steiglitz2009c}. This beautiful idea did not attract much attention because a realistic device, which enables the all-optical transportation of a relatively weak state of light including its loading and unloading, has not been suggested to date.

In this paper, we describe a microdevice where relatively weak optical pulses and eigenstates are transported between input and output ports by a soliton (Fig. \ref{fig:fig2}). It consists of an uncoated optical fiber segment (FS) coupled to three transverse input-output microfibers, $\mathrm{MF}_0$, $\mathrm{MF}_1$ and $\mathrm{MF}_2$. In our model, a whispering gallery soliton (WGS) is formed by resonant excitation of a whispering gallery mode launched by the $\mathrm{MF}_0$ inside the FS. The WGS slowly propagates along the surface of the fiber and further slows down near $\mathrm{MF}_1$ where a relatively weak whispering gallery pulse or eigenstate (WGE) is loaded. Similar to the slow linear propagation of whispering gallery modes realized in SNAP technology \cite{Sumetsky2013b, Hamidfar2018a, Sumetsky2012a, Toropov2016a}, we engineer the nanoscale variation of the fiber effective radius (corresponding to the sub-GHz variation of the cutoff frequency) so that the soliton can slow down, stop, as well as reverse its propagation direction. In particular, the WGS can continue its propagation to $\mathrm{MF}_2$ where the WGE is unloaded, or turn back before reaching $\mathrm{MF}_2$ and unload the WGE back into $\mathrm{MF}_1$. Furthermore, we suggest that the required dramatically small variation of the fiber parameters along its length can be introduced all-optically.

A pulse with central angular frequency $\omega_s$ is coupled into the FS from $\mathrm{MF}_0$ forming a whispering gallery mode, which is enhanced due to the constructive self-interference. As a result, a WGS with central frequency $\omega_s$ is formed. It is assumed that $\omega_s$ is close to the cutoff frequency $\omega_s^{\mathrm{(cut)}}(z)$ of the FS, which is slowly varying along the FS axis $z$. Due to the proximity of $\omega_s$ and $\omega_s^{\mathrm{(cut)}}(z)$, the axial speed of the created WGS and is small and, for this reason, sensitive to a small variation of $\omega_s^{\mathrm{(cut)}}(z)$. Similarly, the central frequency of a weak WGE, $\omega_e$, different from  $\omega_s$, is close to a slowly and weakly varying cutoff frequency $\omega_e^{\mathrm{(cut)}}(z)$. Due to the small cutoff frequency variations $\Delta\omega_{s,e}^{\mathrm{(cut)}}(z) = \omega_{s,e}^{\mathrm{(cut)}}(z) - \omega_{s,e}$ assumed here, the expression for slowly propagating whispering gallery modes can be factorized as $R_{m_{s,e}p_{s,e}}(r)e^{im_{s,e}\varphi }{\psi }_{s,e}(z,t)e^{-i\omega_{s,e}t}$ where  $m_{s,e}$ and $p_{s,e}$ are azimuthal and radial quantum numbers. Consequently, the propagation of a narrow bandwidth WGS with central frequency $\omega_s$ and a WGE with frequency $\omega_e \neq \omega_s$ along the fiber axis coordinate $z$ is determined by their amplitudes $\psi_s(z)$ and $\psi_e(z)$. These functions are defined by a system of coupled nonlinear Schr\"{o}dinger equations, which are similar to those commonly used in nonlinear fiber optics \cite{Agrawal2012, Manassah1990a, Steiglitz2009c} where the temporal and spatial coordinates are interchanged \cite{DeSterke1988,Bhat2001,Weiss2018, Suchkov2017}. Assuming $|\psi_s(z)|\gg |\psi_e(z)|$ we have:

\begin{widetext}
\begin{subequations}\label{eq:eq1}
\begin{align}
& i\partial_t \psi_s = -\frac{1}{2}\kappa_s\partial_z^2\psi_s + \bigg[ \Delta\omega_s^{\mathrm{(cut)}}(z) + i\gamma_s + \kappa_s\sum_{j=1}^2 D_{sj}\delta(z-z_j) - \frac{\omega_s n_2}{n_s A_{ss}} |\psi_s|^2 \bigg]\psi_s + J_s(t)\delta(z-z_0) \label{eq:sub1a}\\
& i\partial_t \psi_e = -\frac{1}{2}\kappa_e\partial_z^2\psi_e + \bigg[ \Delta\omega_e^{\mathrm{(cut)}}(z) +i\gamma_e + \kappa_e\sum_{j=1}^2 D_{ej}\delta(z-z_j) - 2\frac{\omega_e n_2}{n_e A_{se}} |\psi_s|^2 \bigg]\psi_e + J_e(t)\delta(z-z_1) \label{eq:sub1b}
\end{align}
\end{subequations}
\end{widetext}

Here $\kappa_{s,e}= c^2/(n_{s,e}^2\omega_{s,e})$, $c$ is the speed of light, $n_s$ and $n_e$ are the refractive indices of the FS at frequencies $\omega_s$ and $\omega_e$, $n_2$ is its nonlinear refractive index, $\delta(z)$ is the delta-function, and $A_{ss}$, $A_{se}$ are the effective mode areas defined in \cite{Agrawal2012} and Appendix A1. The terms $J_s(t)\delta(z-z_0)$ and $J_e(t)\delta(z-z_1)$ in Eqs. \eqref{eq:sub1a} and  \eqref{eq:sub1b} determine the soliton and weak pulse sources at microfibers, which are specified below. Parameters $D_{sj}$ and $D_{ej}$ are the coupling of the FS to microfibers at frequencies $\omega_s$ and $\omega_e$ determined in Ref. \cite{Sumetsky2012a}. For a single input-output microfiber with a source, Eq. \eqref{eq:sub1a} coincides with that obtained previously in \cite{Suchkov2017}.

To estimate the characteristic parameters of our device, we assume that the FS is uniform. Then Eqs. \eqref{eq:sub1a} and \eqref{eq:sub1b} can be solved analytically \cite{Agrawal2012} yielding for WGS: 

\begin{equation}\label{eq:eq2}
| \psi_s^{(0)}(z,t) |^2 = P_s \sech^2{\bigg(\frac{z-v_st}{L_s}\bigg)}, \ P_s = \frac{c^2A_{ss}}{n_s n_2 \omega_s^2 L_s^2},
\end{equation}

\noindent where $P_s$ is the soliton peak power, $L_s$ is the soliton characteristic width and $v_s$ is the soliton velocity. After the substitution of $|\psi_s^{(0)}(z,t) |^2$ from Eq. \eqref{eq:eq2}, Eq. \eqref{eq:sub1b} describes the propagation of the WGE along the FS with time-dependent cutoff frequency $\omega_e^{\mathrm{(cut)}}(z) - \Delta\omega_e^{\mathrm{max}}\sech^2{\big(\frac{z-v_st}{L_s}\big)}$ where

\begin{equation}\label{eq:eq3}
\Delta \omega_e^{max} = \frac{2c^2 \omega_e A_{ss}}{n_e n_s \omega_s^2 L_s^2A_{se}}= 2 P_s \frac{\omega_e n_2}{n_e A_{se}}.
\end{equation}

Eq. \eqref{eq:eq3} determines the maximum variation of the cutoff frequency caused by the WGS. Assuming that the WGE have the same speed as the WGS, we look for the solution of Eq. \eqref{eq:sub1b} in the form $\psi^{(0)}_e(z,t) = \Phi(x)e^{i\alpha x + i\beta t}$, which depends on the dimensionless relative coordinate $x = \frac{z - v_s t}{L_s}$. Then $\Phi(x)$ satisfies the equation

\begin{equation}\label{eq:eq4}
\frac{d^2 \Phi(x)}{d x^2} + \big( \epsilon - \eta \sech^2{x} \big)\Phi(x) = 0, \ \eta = \frac{2n_e\omega_e^2A_{ss}}{n_s\omega_s^2A_{se}},
\end{equation}

\noindent where $\epsilon = \frac{L_s^2}{c^4}\big[  2n_e^2\omega_e c^2 (\Delta\omega_e + \beta) - v_s^2n_e^4\omega_e^2  \big]$. Parameter $\epsilon$ can be tuned by varying $\Delta\omega_e = \omega_e^{(cut)} - \omega_e$. The eigenvalues $\epsilon_m$ of Eq. \eqref{eq:eq4} corresponding to the localized states $\Phi_m(x)$ are 

\begin{equation}\label{eq:eq5}
\epsilon_m = -(\xi - m)^2, \quad \xi = \sqrt{\eta + 1/4} - 1/2
\end{equation}

\noindent and the index $m = 0,1,...,[\xi]$. The integer part of $\xi$, $[\xi]$, is the total number of localized eigenstates that the WGS can support. If $\eta < 2$ then Eq. \eqref{eq:eq4} has a single eigenstate $\Phi_0(x) = \sech^{\xi }x$. For small values of $\eta $ we have $\xi \approx 2\eta $, $\epsilon_0 \approx  4 \eta^2$ , and the characteristic width of this eigenstate is $x_w = 1/\eta $ which corresponds to the WGE axial width $z_w = L_s / \eta $.


\begin{figure*}[htbp]
\includegraphics[width=\linewidth]{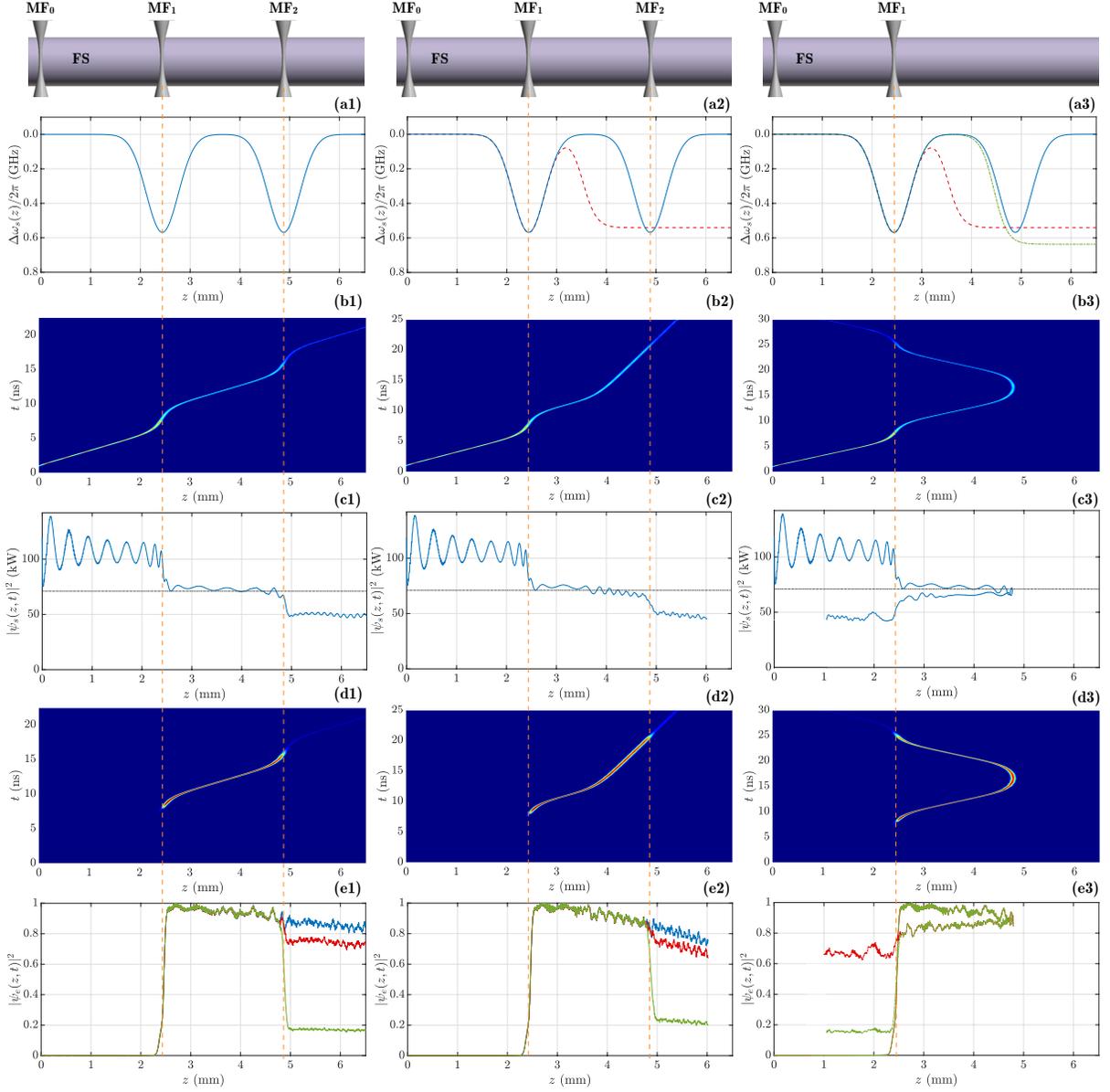}%
\caption{Transportation of a WGE by a WGS. (a1), (a2), and (a3) Variation in the cutoff frequency introduced to control the speed of the WGS in the three cases considered (solid blue, dashed red, and dash-dotted green curves). (b1), (b2) and (b3) Evolution of WGS. (c1), (c2) and (c3) Variation of the WGS peak power as it propagates along the fiber. (d1), (d2) and (d3) Evolution of WGE positioned at the ground state of the WGS-induced quantum well. (e1), (e2) and (e3) Variation of the maximum of the WGE as it propagates along the fiber for different coupling coefficients with microfiber $\mathrm{MF}_2$ (red curves: $D_{s2} = 0.005i \ \mu \mathrm{m}^{-1}$ and $D_{e2} = 0$; blue curves: $D_{s2} = 0$ and $D_{e2} = 0.05 \ \mu \mathrm{m}^{-1}$; green curves: $D_{s2} = 0.005i \ \mu \mathrm{m}^{-1}$ and $D_{e2} = 0.05i \ \mu \mathrm{m}^{-1}$)} 
\label{fig:fig3}
\end{figure*}


Fig.\ref{fig:fig3} shows three exemplary voyages of a WGE transported by a WGS along a silica FS with radius $r_0=20 \ \mu \mathrm{m}$ found by numerical solution of Eqs. \eqref{eq:sub1a} and \eqref{eq:sub1b} (see Appendix A1). This figure includes three examples with the cutoff frequency variations shown in Fig. \ref{fig:fig3}(a1), (a2), and (a3), corresponding to acceleration, slowing down, and stopping of WGS between $\mathrm{MF}_1$ and $\mathrm{MF}_2$. We design the FS profile to satisfy the condition of adiabaticity, which ensures that the shapes of WGS and WGE are not altered significantly during the propagation. In addition, the speed of the WGS near $\mathrm{MF}_1$ and $\mathrm{MF}_2$ is set to enable loading, safe transporting and unloading of the WGE. The central frequency of the input source in Eq. \eqref{eq:sub1a} generating the WGS at $\mathrm{MF}_0$  is set to $\omega_s/2\pi =225 \ \mathrm{THz}$. In order to arrive at the minimum possible WGS speed, this frequency is assumed to coincide with the value of the FS cutoff frequency at $\mathrm{MF}_0$. The characteristic width, speed, and duration of the created soliton is  $L_s \sim 50 \ \mu \mathrm{m}$, $v_s \sim 0.5 \ \mathrm{mm/ns}$ and $T_s \sim 100\ \mathrm{ps}$. The frequency of the plane wave entering the FS from $\mathrm{MF}_1$ and forming the WGE is set to $195 \ \mathrm{THz}$ plus a small shift to match one of the possible eigenfrequencies of the WGS-formed potential well.  Attenuations of WGS and WGE are set to $\gamma_{s,e} = 3 \ \mathrm{MHz}$ (corresponding to a Q-factor $\sim 2 \cdot 10^8$ at frequency $\omega/2\pi = 190 \ \mathrm{THz}$ \cite{Gorodetsky1996, Pollinger2009a}. Other parameters of our devices are described in the Appendix A1. The height of the potential well formed by the WGS found from Eq. \eqref{eq:eq3} is $\Delta\omega^{\mathrm{max}}_e/2\pi \sim 2.6\ \mathrm{GHz}$, while the WGS maximum power is $P_s \sim 70 \ \mathrm{kW}$. From Eq. \eqref{eq:eq5}, this potential well can only support a single WGE. For the WGS to survive and in order to minimize the WGE perturbation during its loading and unloading, it is critical to minimize the WGS coupling to $\mathrm{MF}_1$ and $\mathrm{MF}_2$ while keeping the WGE coupling large. This can be achieved by appropriate phase-matching microfiber-WGE coupling at frequency $\omega_e$ and phase unmatching of microfiber-WGS coupling at frequency $\omega_s$ \cite{Little1999}. In our modeling, we set $D_{s1} = D_{s2} = 0.005i\ \mu \mathrm{m}^{-1}$ and $D_{e1}=D_{e2}=0.05i\ \mu \mathrm{m}^{-1}$, where the latter correspond to characteristic experimentally observed values \cite{Sumetsky2012a,Vitullo2020}. We found that the real parts of these coupling parameters with the same order of magnitude do not noticeably modify the WGE (blue curve in Fig. \ref{fig:fig3}(e1), (e2)). This is the reason why they are set to zero.

Figs. \ref{fig:fig3}(b1)-(b3) show the propagation of the WGS considered in the examples. In Fig. \ref{fig:fig3}(b1), the WGS speed in the intervals between input-output microfibers is $4.7\cdot 10^5 \ \mathrm{m/s} = 0.0016 c$ and decreases to $0.9\cdot 10^5 \ \mathrm{m/s} = 0.0003c$ near $\mathrm{MF}_1$ and $\mathrm{MF}_2$ for loading and unloading the WGE. Fig. \ref{fig:fig3}(b2) shows the propagation of the WGS when its speed is reduced to $1.3\cdot 10^5\ \mathrm{m/s}= 0.00044c$ in between microfibers. In Fig. \ref{fig:fig3}(b3), the WGS stops in between $\mathrm{MF}_1$ and $\mathrm{MF}_2$ and returns back to $\mathrm{MF}_1$ to unload the WGE (see \cite{Sumetsky2013b} for the analogue dispersionless propagation of a linear pulse). Figs. \ref{fig:fig3}(c1)-(c3) show the variation of the WGS power during its propagation. In order not to destroy the WGS due to its leakage through $\mathrm{MF}_1$ and form an almost pure first order soliton ideal for transportation of signals between $\mathrm{MF}_1$ and $\mathrm{MF}_2$, the WGS original power was made high enough. Finally, Figs. \ref{fig:fig3}(d1)-(d3) and Figs. \ref{fig:fig3}(e1)-(e3) describe the loading, transport, and unloading of the WGE corresponding to the ground eigenstate of the WGS-formed quantum well. In order to load this eigenstate into the WGS-formed quantum well, we tuned the frequency of the input wave to match the frequency of this eigenstate. The detuning of the ground WGE eigenfrequency $\delta \omega /2\pi $ from the FS cutoff frequency is $0.88 \ \mathrm{GHz}$. Notice that a relatively small coupling $D_{s2}$ still contributes to the WGE dissipation into the FS. The latter is found by setting $D_{e2}=0$ (red curves in Fig. \ref{fig:fig3}(e1)-(e3)). Taking this dissipation into account, we find that, for the $D_{e2}= i0.05 \ \mu \mathrm{m}^{-1}$ chosen, more than $70\%$ of the WGE power is unloaded into $\mathrm{MF}_2$. Obviously, the amount of unloaded power can be improved by increasing $D_{e2}$. Generally, the evolution of a WGS-trapped WGE pulse with frequencies distributed within the quantum well bandwidth (rather than coinciding with its eigenvalue) can be quite complex \cite{Schrodinger1926, Sumetsky2015a, Robinett2004}. 

Thus, as in the case of slow linear propagation of whispering gallery modes \cite{Sumetsky2013b, Hamidfar2018a, Sumetsky2012a, Toropov2016a}, the speed of a slow WGS can be controlled by the dramatically small variation of cutoff frequency along the optical fiber length. Such WGS can serve as a moving optical microresonator -- a soliton micro-truck -- enabling programmed transportation of optical pulses and eigenstates including their loading and unloading. Various other examples of transportation of a weak signal by solitons as well as by optical pulses having power below the soliton formation threshold can be considered. Of special interest is the investigation of nonadiabatic processes during the transportation of a WGE \cite{Miyashita2007}, its loading, and unloading.  The characteristic peak power and duration of a WGS propagating along a silica fiber considered here, are $100$ kW and $100$ ps, respectively. Such strong pulses may result in fiber damage (see Appendix A3 and Refs. \cite{Stuart1995a,Smith2009}). Therefore, further optimization of the carrier pulse parameters and its speed may be required. However, according to our estimates (see Appendix A3), these pulses do not introduce a significant temperature variation. For chalcogenide and hydrogenated amorphous silicon (a-Si:H) fibers, which have larger nonlinearity, the peak power of the WGS can be two orders of magnitude smaller \cite{Shiryaev2017, Vukovic2013a}. As it is well-known from quantum mechanics \cite{Landau1981}, the one-dimensional potential well induced by a WGS can always hold at least one optical eigenstate despite of its shallowness. One of the intriguing conclusion of our findings is that the WGS speed can be fully controlled by unexpectedly small variation of the cutoff wavelength $\Delta\omega_s^{\mathrm{(cut)}} \sim 1\ \mathrm{GHz}$, which, for the fiber radius $r_0\sim 20\ \mu \mathrm{m}$, corresponds to an effective radius variation of $r_0 \Delta\omega_s^{\mathrm{(cut)}}/\omega_s\sim \ 0.1\ nm$. The fabrication precision achievable in SNAP technology \cite{Sumetsky2013b,Toropov2016a} makes the introduction of such dramatically small variations feasible. Furthermore, these variations can be induced all-optically. In fact, the amplitude of mechanical vibrations of an optical microresonator, which are induced by whispering gallery modes, can be tuned up to $10$ nm \cite{Anetsberger2009}. For the microresonator with radius $r_0 \sim 20\ \mu \mathrm{m}$ considered in \cite{Anetsberger2009}, this corresponds to a cutoff frequency variation exceeding $10$ GHz. Launched through the same or additional control input-output microfibers, these modes can temporary induce the required variation of the cutoff wavelength. In this case, the behavior of WGS (or a carrier pulse with power below the soliton threshold) and WGE are determined by the same pair of Eqs. \eqref{eq:eq1} where the cutoff frequency variations $\Delta \omega_{s,e}^{\mathrm{(cut)}}(z,t)$ now depend on time $t$. Thus, the device described here potentially enables the all-optically controlled transportation of light by light at the microscale.

\bibliography{NLSNAP}

\renewcommand{\theequation}{A\arabic{equation}}
\renewcommand{\thefigure}{A\arabic{figure}}

\setcounter{equation}{0}
\setcounter{figure}{0}

\section*{\Large{Appendix}}

\section*{A1. Numerical solution of nonlinear Schr\"{o}dinger equations}\label{sec:secA1}

We consider the nonlinear Schr\"{o}dinger equations (1a) and (1b) 

\begin{subequations}\label{eq:eqS1}
\begin{align}
& i\partial_t \psi_s = -\frac{1}{2} \kappa_s \partial_z^2\psi_s - \frac{\omega_s n_2}{n_sA_{ss}} |\psi_s|^2 \psi_s +\bigg[ \Delta\omega_s(z)+i\gamma_s + \kappa_s\sum_{j=1}^2 D_{s;j}\delta(z-z_j) \bigg]\psi_s + J_s(t)\delta(z-z_0) \label{eq:subS1a}\\
& i\partial_t \psi_e = -\frac{1}{2} \kappa_e \partial_z^2\psi_e - 2\frac{\omega_e n_2}{n_e A_{se}} |\psi_s|^2 \psi_e +\bigg[ \Delta\omega_e(z)+i\gamma_e + \kappa_e\sum_{j=1}^2 D_{e;j}\delta(z-z_j) \bigg]\psi_e + J_e(t)\delta(z-z_1)\label{eq:subS1b}
\end{align}
\end{subequations}

In the equations, $\Delta\omega_{s,e}(z) = \omega_{s,e}^{\textrm{(cut)}}(z)-\omega_{s,e}$ is the cutoff frequency variation and

\begin{equation}\label{eq:eqS2}
\kappa_{s,e} = \frac{c^2}{\omega_{s,e} n_{s,e}^2}.
\end{equation}

The subscripts $s,e$ refer to the WGS and the WGE respectively. The values of the parameters used in our numerical simulations are defined as follows. The speed of light $c=3\cdot{10}^8$ m/s, the WGS central frequency $\omega_s/2\pi =225$ THz, and the WGE frequency $\omega_e/2\pi =195$ THz. For the silica fiber considered, we set the refractive indices at these frequencies equal to $n_s=n_e=1.47$. The nonlinear refractive index is $n_2=2.5\cdot {10}^{-8}\ \mu \mathrm{m}^2/\mathrm{W}$. The terms $\gamma_{s,e}$ are the attenuation constants determined by the losses of our system; their values are set to $\gamma_s=\gamma_e=3\ \mathrm{MHz}$, which corresponds to a Q factor of $\sim 2 \cdot 10^8$. The coupling coefficients are set to $D_{s,1}=D_{s,2}=i0.005\ \mu \mathrm{m}^{-1}$ for the WGS and $D_{e,1}=D_{e,2}=i0.05\ \mu \mathrm{m}^{-1}$ for the WGE. The effective mode areas in Eqs. \eqref{eq:eqS1} are given by

\begin{equation}\label{eq:eqS3}
\begin{aligned}
& A_{se} = \frac{\Big( \iint dS\big| F_s(r,\phi) \big|^2 \Big)\Big( \iint dS\big| F_e(r,\phi) \big|^2 \Big)}{\iint dS\big| F_s(r,\phi) \big|^2\big| F_e(r,\phi) \big|^2} \\
& A_{ss} = \frac{\Big[ \iint dS\big| F_s(r,\phi) \big|^2 \Big]^2}{\iint dS\big| F_s(r,\phi) \big|^4}
\end{aligned}
\end{equation}

\noindent where $F_{s,e}(r,\phi )$ are the transversal modal distribution of WGS and WGE [1]. In order to calculate the effective mode areas, we need to evaluate these integrals. The transversal modal distribution are approximated by the Airy functions:

\begin{equation}\label{eq:eqS4}
F_{s,e}(r,\phi )\cong e^{im_{s,e}\phi }\ \mathrm{Ai}[(2m^2_{s,e})^{1/3}(1-r/r_0)+{\zeta }_i]
\end{equation}

Here azimuthal quantum numbers $m_{s,e}$ are related to the frequencies and the fiber radius $r_0$ by $m_{s,e}=\omega_{s,e}n_{s,e}r_0/c$ and ${\zeta }_i$ are the zeros of the Airy function. For the particular frequencies used and a fiber radius $r_0=20\ \mu \mathrm{m}$, and the first radial mode with $\zeta_1=-2.338$, the effective mode areas are $A_{ss}=170.7\ \mu \mathrm{m}^2$ and $A_{se}=161.8\ \mu \mathrm{m}^2$.

We approximate the delta functions in Eqs. \eqref{eq:eqS1} by the functions $\frac{1}{\pi d}{\mathrm{sech} \left[(z-z_j)/d\right]\ }$ with $d=2.5\ \mu \mathrm{m}$. Here $d$ determines the characteristic size of the microfiber-FS coupling. We have chosen a Gaussian pulse as the source $J_s(t)$. In particular, for our simulations, $J_s(t)=\mathrm{exp}[-(t-t_0)^2/\tau_0^2]$ where $t_0=1$ ns and $\tau_0=0.3$ ns. The weak signal source $J_e(t)$ corresponds to a CW signal with frequency $\omega_e+\delta\omega_e$, where $\delta\omega_e$ is the detuning chosen so that  $\omega_e+\delta\omega_e$ is equal to the eigenfrequency of the WGS potential.

In our numerical simulations, we use the dimensionless version of Eqs. \eqref{eq:eqS1} by introducing dimensionless variables

\begin{equation}\label{eq:eqS5}
\begin{aligned}
\tau = \frac{t}{T_0}; \quad \xi = \frac{z}{L_0}; \quad \hat{\psi}_{s,e} = \frac{\psi_{s,e}}{\sqrt{P_0}}; 
\end{aligned}
\end{equation}

\noindent which gives

\begin{subequations}\label{eq:eqS6}
\begin{align}
i\partial_{\tau} \hat{\psi}_s & = -\frac{1}{2} \frac{\kappa_sT_0}{L_0^2} \partial_{\xi}^2\hat{\psi}_s - \frac{\omega_s n_2 T_0 P_0}{n_sA_{ss}}|\hat{\psi}_s|^2 \hat{\psi}_s + \notag \\
& + T_0\bigg[ \Delta\omega_s(z)+i\gamma_s + L_0\sum_{j=1}^2 D_{s;j}\delta_j(\xi-\xi_j) \bigg]\hat{\psi}_s + \frac{T_0}{L_0\sqrt{P_0}}J_s(\tau)\delta_0(\xi-\xi_0) \label{eq:subS6a} \\
i\partial_{\tau} \hat{\psi}_e & = -\frac{1}{2} \frac{\kappa_eT_0}{L_0^2}\partial_{\xi}^2\hat{\psi}_e - 2\frac{\omega_e n_2T_0 P_0}{n_eA_{se}}|\hat{\psi}_s|^2 \hat{\psi}_e + \notag \\
&+ T_0\bigg[ \Delta\omega_e(z)+i\gamma_e + r_{\omega}^{-1}L_0\sum_{j=1}^2 D_{e;j}\delta_j(\xi-\xi_j) \bigg]\hat{\psi}_e + \frac{T_0}{L_0\sqrt{P_0}}J_e(\tau)\delta_1(\xi-\xi_1) \label{eq:subS6b}
\end{align}
\end{subequations}

We choose time $T_0$ by setting $\frac{\kappa_s T_0}{L_0^2}=1$ so that

\begin{equation}\label{eq:eqS7}
\begin{aligned}
L_0 = \sqrt{k_sT_0}
\end{aligned}
\end{equation}

In addition, we choose $P_0$ by setting $\frac{\omega_s n_2 T_0 P_0}{n_sA_{ss}} = 1$ which gives

\begin{equation}\label{eq:eqS8}
\begin{aligned}
& P_0  = \frac{c^2 A_{ss}}{\omega_s^2n_sn_2L_0^2}
\end{aligned}
\end{equation}

Finally, we introduce the following dimensionless parameters and functions:

\begin{equation}\label{eq:eqS9}
\begin{aligned}
\Delta\hat{\omega}_{s,e} = \Delta\omega_{s,e}\cdot T_0; \quad \hat{\gamma}_{s,e} = \gamma_{s,e}\cdot T_0; \quad \hat{D}_{s} = D_{s}\cdot L_0; \quad \hat{D}_{e} = D_{e}\cdot r_{\omega}^{-1}L_0; \quad \hat{J}_{s,e} = J_{s,e}\cdot \frac{T_0}{L_0 \sqrt{P_0}}
\end{aligned}
\end{equation}

As the result, Eqs. \eqref{eq:eqS6} are presented in the dimensionless form:

\begin{subequations}\label{eq:eqS10}
\begin{align}
& i\partial_{\tau} \hat{\psi}_s = -\frac{1}{2} \partial_{\xi}^2\hat{\psi}_s - |\hat{\psi}_s|^2 \hat{\psi}_s + \bigg[ \Delta\hat{\omega}_s+i\hat{\gamma}_s + \sum_{j=1}^2 \hat{D}_{s;j}\delta_j(\xi-\xi_j) \bigg]\hat{\psi}_s + \hat{J}_s(\tau)\delta_0(\xi-\xi_0) \label{eq:subS10a}\\
& i\partial_{\tau} \hat{\psi}_e = -\frac{1}{2} r_{\omega}^{-1}\partial_{\xi}^2\hat{\psi}_e - r_{\omega}^{-1}\eta|\hat{\psi}_s|^2 \hat{\psi}_e + \bigg[ \Delta\hat{\omega}_e+i\hat{\gamma}_e + \sum_{j=1}^2 \hat{D}_{e;j}\delta_j(\xi-\xi_j) \bigg]\hat{\psi}_e + \hat{J}_e(\tau)\delta_1(\xi-\xi_1)\label{eq:subS10b}
\end{align}
\end{subequations}

\noindent where $r_{\omega} = \omega_e/\omega_s = 0.87$ is the ratio of frequencies,

\begin{equation}\label{eq:eqS11}
\begin{aligned}
\eta = \frac{2n_e\omega_e^2A_{ss}}{n_s\omega_s^2A_{se}} = 2 r_n r_{\omega}r_{A}
\end{aligned}
\end{equation}

\noindent is the dimensionless height of the soliton potential (see Eq. \eqref{eq:eq4}) where  $r_A=A_{ss}/A_{se} \sim 0.95$ is the ratio of effective mode areas and $r_n=\frac{n_e}{n_s}=1$ is the ratio of refractive indices. For these parameters, the dimensional height of the WGS-induced potential is $\eta =1.435$.

Under quite general conditions [2], we assume that the relative cutoff frequency variations of the FS at the WGS and WGE frequencies are equal:

\begin{equation}\label{eq:eqS12}
\begin{aligned}
\frac{\Delta\omega_s(z)}{\omega_s} = \frac{\Delta \omega_e(z)}{\omega_e}
\end{aligned}
\end{equation}

Taking this relation into account, we finally obtain:

\begin{subequations}\label{eq:eqS13}
\begin{align}
& i\partial_{\tau} \hat{\psi}_s = -\frac{1}{2} \partial_{\xi}^2\hat{\psi}_s - |\hat{\psi}_s|^2 \hat{\psi}_s + \bigg[ \Delta\hat{\omega}_s(\xi)+i\hat{\gamma}_s + \sum_{j=1}^2 \hat{D}_{s;j}\delta_j(\xi-\xi_j) \bigg]\hat{\psi}_s + \hat{J}_s(\tau)\delta_0(\xi-\xi_0) \label{eq:subS13a}\\
& i\partial_{\tau} \hat{\psi}_e = -\frac{1}{2} r_{\omega}^{-1}\partial_{\xi}^2\hat{\psi}_e - r_{\omega}^{-1}\eta|\hat{\psi}_s|^2 \hat{\psi}_e + \bigg[ r_{\omega}\Delta\hat{\omega}_s(\xi)+i\hat{\gamma}_e + \sum_{j=1}^2 \hat{D}_{e;j}\delta_j(\xi-\xi_j) \bigg]\hat{\psi}_e + \hat{J}_e(\tau)\delta_1(\xi-\xi_1)\label{eq:subS13b}
\end{align}
\end{subequations}

We solve Eqs. \eqref{eq:eqS13} numerically using the split-step Fourier method with a uniform grid of $2^{13}$ points in the time window of length $p \cdot T_0$, where $p$ is an integer that depends on the particular simulation scenario and  $T_0=100\ ps$. For example, $p=200$ for the transportation of the WGE in Figs. 3 (a1)-(e1) of the main text. 

\section*{A2. Effect of propagation losses and comparison with a weaker carrier pulse}\label{sec:secA2}

\begin{figure}[h!]
\begin{center}
\includegraphics[width=0.45\textwidth]{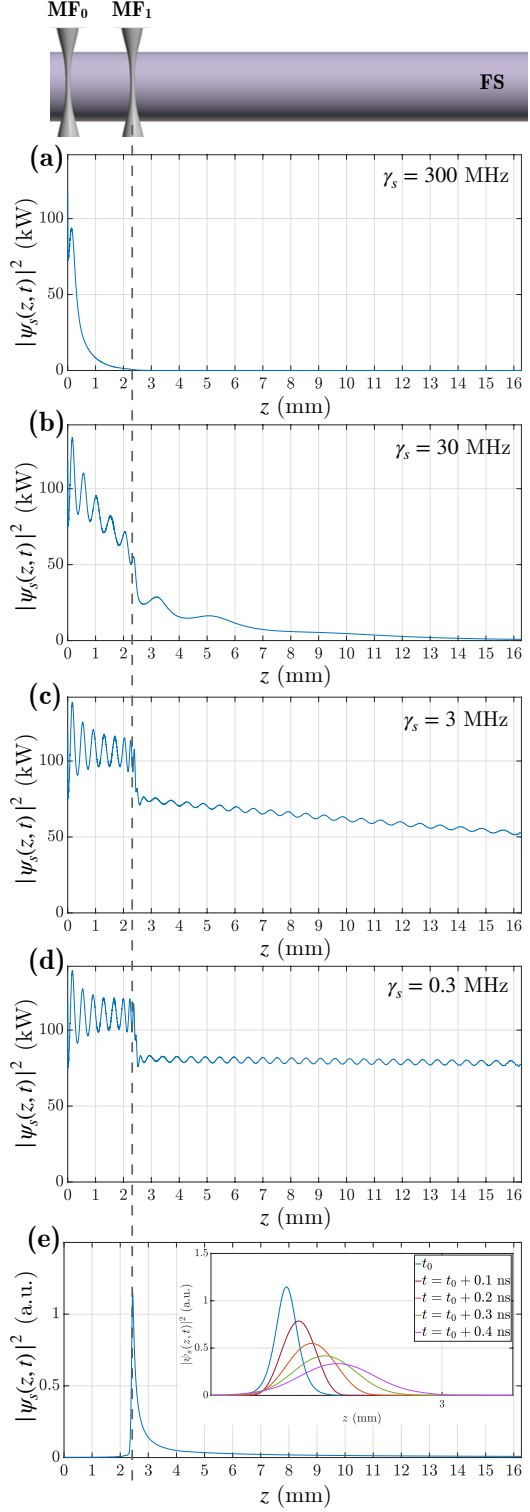}
\caption[1\textwidth]{Effect of losses on the WGS propagation for different attenuation constants: (a) $\gamma_s=300 \ \mathrm{MHz}$. (b) $\gamma_s=30 \ \mathrm{MHz}$. (c) $\gamma_s=3 \ \mathrm{MHz}$. (d) $\gamma_s=0.3 \ \mathrm{MHz}$. (e) Linear and lossless propagation of a WGS. The inset shows the evolution of the soliton profile as it disperses.}
\label{fig:figS1}
\end{center}
\end{figure}

Traveling between stops, WGS and WGE experience absorption and scattering losses. The effect of these losses is characterized by the attenuation factors $\gamma_{s,e}$. The distance that the WGS and the WGE can travel without significant attenuation also depends on the speed of the WGS, which is determined by the shape of the source in Eq. \eqref{eq:subS1a} and the deviation of the frequencies $\omega_{s,e}$ from the corresponding cutoff frequencies. Here, we investigate the effect of losses by considering the propagation of a WGS in a microdevice with two microfibers, $\mathrm{MF}_0$ and $\mathrm{MF}_1$, having the coupling parameters indicated in Appendix A1. A WGS launched from $\mathrm{MF}_0$ passes  $\mathrm{MF}_1$ and then propagates along the uniform part of the FS. Fig. \ref{fig:figS1}(a) shows the evolution of the WGS with an attenuation factor of $\gamma_s=300$ MHz (corresponding to $Q=2\times 10^6$ at frequency $\omega/2\pi =190\ \mathrm{THz}$). The plot shows that the WGS is not fully formed and does not reach $\mathrm{MF}_1$. If we reduce the attenuation factor to $\gamma_s=30\ \mathrm{MHz}$ ($Q=2\times 10^7$), the WGS is formed as shown in Fig. \ref{fig:figS1}(b). However, the soliton quickly decays, in particular, after passing $\mathrm{MF}_1$ where it experiences additional losses of energy due to the coupling with $\mathrm{MF}_1$. In Fig. \ref{fig:figS1}(c),   $\gamma_s=3\ \mathrm{MHz}$ ($Q=2\times 10^8$). In this case, which has been considered in the main text, a WGE can be formed and survive a few millimeters of transportation. In Fig. \ref{fig:figS1}(d),  the attenuation is reduced further to $\gamma_s=0.3\ \mathrm{MHz}$ ($Q=2\times 10^9$). This value for the attenuation factor is feasible [3] and allows our microdevice to transport the WGE from one microfiber to the other without significant losses. Finally, for comparison, Fig. \ref{fig:figS1}(e) shows the linear and lossless propagation of a pulse launched from the vicinity of $\mathrm{MF}_1$, which has the same original shape as the WGS which has just passed $\mathrm{MF}_1$ in Fig. \ref{fig:figS1}(d). The propagation of this pulse was calculated by setting $\gamma_s=0$ and $n_2=0$ in Eq. \eqref{eq:subS1a}. It can be seen that, due to dispersion, this pulse strongly decays before reaching $\mathrm{MF}_1$. In addition, as it follows from the inset in Fig. \ref{fig:figS1}(e), the speed of the pulse spreading is comparable with its group velocity.

\section*{A3. The damage threshold and temperature effects}\label{sec:secA3}

The damage on the optical fiber induced by the CW radiation is usually due to melting as a result of light energy absorption. However, pulses shorter than $ \sim 1$ ns can damage the optical fiber by other processes including dielectric breakdown in the material (electron avalanche) caused by the strong electric field (see [4-6] and references therein). The threshold for optical damage of short pulses depends on the pulse central frequency and its intensity distribution in space and time. The threshold values experimentally determined previously are characterized by the value of fluence $F$ defined as the average energy of the pulse per its unit cross-sectional area. Roughly assuming that $F \sim \omega^{0.4}$ [4] and using the data of Ref. [5] we find that the threshold fluence for a pulse with duration of $0.1$-$1$ ns is $10$-$50$ $\mathrm{J}/\mathrm{cm}^{2}$. These values might not be directly applicable to the WGS due to the specific geometry of our problem and they can significantly vary depending on experimental conditions. However, we believe that they can serve as reasonable estimates for the problem considered here. 

\begin{figure}[h!]
\begin{center}
\includegraphics[width=1\textwidth]{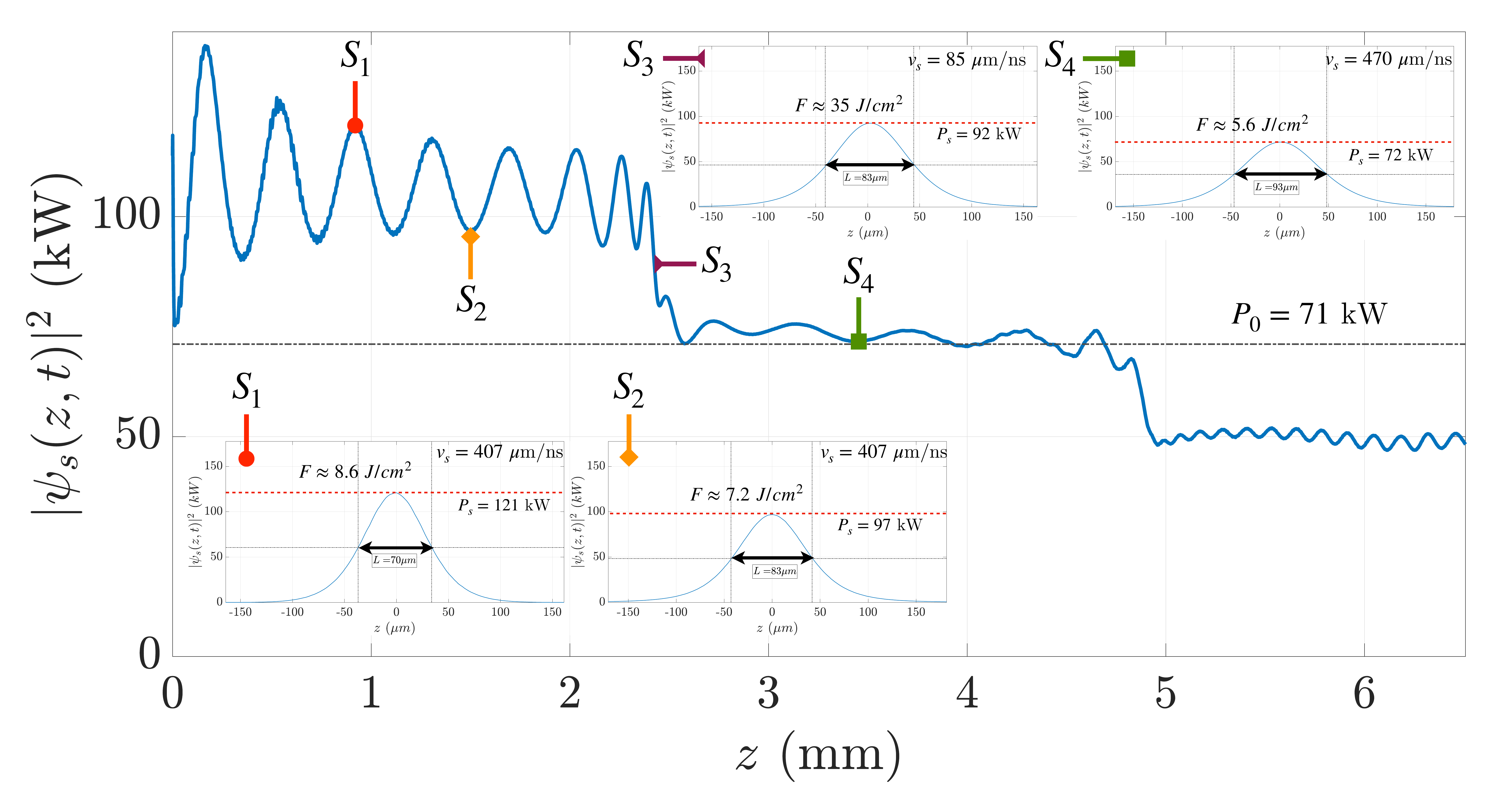}
\caption[0.5\textwidth]{Variation of maximum of the WGS power along the fiber. Insets correspond to the WGS spatial profile at the particular WGS positions indicated on the figure.}
\label{fig:figS2}
\end{center}
\end{figure}

We calculate the fluence $F$ of the WGS using the soliton model described by Eq. (2) of the main text which yields

\begin{equation}\label{eq:eqS14}
\begin{aligned}
F=\frac{1.76c^2}{{\omega }^2_sn_sn_2vL}
\end{aligned}
\end{equation}

Here $L$ and $v$ are the FWHM and the speed of the WGS. In Fig. \ref{fig:figS2}, expanding Fig. 3(c1) of the main text, we show the evolution of the WGS maximum along the FS. The insets show the spatial profile of the soliton at certain positions in between microfibers and during the slowdown at $\mathrm{MF}_1$ indicating its FWHM, maximum power, and speed. The WGS fluences found from Eq. \eqref{eq:eqS14} and indicated in the insets are within the range $7$-$35$ $\mathrm{J}/\mathrm{cm}^2$ corresponding to the possible damage threshold values according to the estimate made above. While further investigation is needed to establish the actual damage threshold of our microdevice, we would like to note here that, as it follows from Eq. \eqref{eq:eqS14}, optimization of the WGS parameters leading to a larger FWHM and speed allows to reduce the value of $F$ significantly.

The propagation of WGS and WGE is controlled by dramatically small variations of the cutoff frequencies along the FS length. Therefore, we have to ensure that these variations are not affected by the temperature variation caused by the WGS propagation or, alternatively, take them into account in our modeling. Attenuation of the WGS power is primary due to absorption and scattering effects, while only the absorption of power contributes to fiber heating. Let us assume that the latter effect is determined by an attenuation coefficient $\gamma_{\mathrm{abs}}=1$ MHz. Since the characteristic cutoff frequency variation and WGS bandwidth $\Delta \omega \sim 1GHz\gg \gamma_{\mathrm{abs}}$, we can estimate the WGS spatial attenuation constant in the linear approximation as (see e.g. [2])

\begin{equation}\label{eq:eqS15}
\begin{aligned}
\alpha =2^{1/2} n_s c^{-1} \omega_s^{1/2} \operatorname{Im} \bigg[ \big(\Delta\omega +i\gamma_{\mathrm{abs}} \big)^{1/2} \bigg] \cong 2^{-1/2} n_{s} c^{-1} \omega_s^{1/2} \Delta\omega^{-1/2} \gamma_{\mathrm{abs}} = 4 \ \mathrm{m}^{-1}
\end{aligned}
\end{equation}

From our numerical simulations shown in Fig. \ref{fig:figS1}(c) and Fig. \ref{fig:figS1}(d), we find $\alpha =7 \ \mathrm{m}^{-1}$  for $\gamma_s=3$ MHz and $\alpha =1.5\ \mathrm{m}^{-1}$  for $\gamma_s=0.3$ MHz. Since only a part of this attenuation contributes to heating, we assume $\alpha =1\ \mathrm{m}^{-1}$. Other parameters of the WGS and FS are set as follows.  Heat capacity and density of silica are $C_p=0.7\ \mathrm{J/g}\cdot \mathrm{K}$ and $\rho =2.65\ \mathrm{g}/\mathrm{cm}^3$; the WGS cross-section, FWHM, speed, and peak power are $A_s=200\ \mu\mathrm{m}^2$, $L=100\ \mu\mathrm{m}$, $v=1\ \mathrm{mm/ns}$, and $P_0=100\ \mathrm{kW}$, respectively. Then, the WGS heating energy is $\Delta E=\frac{P_0L^2}{v}={10}^{-9}\ \mathrm{J}$ and the mass of the FS occupied by the WGS is $m=\rho A_s L =5\cdot {10}^{-8}\ \mathrm{g}$. The corresponding change in temperature is $\Delta T=\frac{\Delta E}{mC_p}=0.03\ \mathrm{K}$. Using the value of $\frac{dn}{dT}=1.3\cdot 10^{-5} \ \mathrm{K}^{-1}$ for silica, we find the shift of the cutoff frequency

\begin{equation}\label{eq:eqS16}
\begin{aligned}
\Delta\omega_T/2\pi = \frac{\omega_s}{2\pi n_s}\frac{d n}{d T}\Delta T= 10^{-2} \ \mathrm{GHz}
\end{aligned}
\end{equation}

This value is noticeably smaller than the characteristic cutoff frequency variation used in our calculations. Generally, the heating effect can be taken into account by modification of the nonlinear Schr\"{o}dinger equations considered here.

\end{document}